\pdfoutput=1

\documentclass[sigconf,screen]{acmart}

\AtBeginDocument{%
  }

\setcopyright{acmcopyright}
\copyrightyear{2023}
\acmYear{2023}
\acmDOI{XXXXXXX.XXXXXXX}

\acmConference[]{}{}{}
\acmPrice{15.00}
\acmISBN{978-1-4503-XXXX-X/18/06}

\usepackage{amsmath,amssymb}  %
\usepackage{multirow}  %
\usepackage{listings}
\usepackage{enumitem}

\begin{document}

\title[CoCon: A Data Set on Combined Contextualized Research Artifact Use]{CoCon: A Data Set on\\Combined Contextualized Research Artifact Use}

\author{Tarek Saier}
\orcid{0000-0001-5028-0109}
\email{tarek.saier@kit.edu}
\affiliation{%
  \institution{Karlsruhe Institute of Technology} %
  \streetaddress{Kaiserstr. 89}
  \city{Karlsruhe}
  \country{Germany}
  \postcode{76133}
}

\author{Youxiang Dong}
\email{youxiang.dong@student.kit.edu}
\affiliation{%
  \institution{Karlsruhe Institute of Technology} %
  \streetaddress{Kaiserstr. 89}
  \city{Karlsruhe}
  \country{Germany}
  \postcode{76133}
}

\author{Michael F{\"a}rber}
\orcid{0000-0001-5458-8645}
\email{michael.faerber@kit.edu}
\affiliation{%
  \institution{Karlsruhe Institute of Technology} %
  \streetaddress{Kaiserstr. 89}
  \city{Karlsruhe}
  \country{Germany}
  \postcode{76133}
}

\renewcommand{\shortauthors}{Saier et al.}

\begin{abstract}
In the wake of information overload in academia, methodologies and systems for search, recommendation, and prediction to aid researchers in identifying relevant research are actively studied and developed. Existing work, however, is limited in terms of granularity, focusing only on the level of papers or a single type of artifact, such as data sets. To enable more holistic analyses and systems dealing with academic publications and their content, we propose $\mathrm{CoCon}$, a large scholarly data set reflecting the combined use of research artifacts, contextualized in academic publications' full-text. Our data set comprises 35 k artifacts (data sets, methods, models, and tasks) and 340 k publications. We additionally formalize a link prediction task for ``combined research artifact use prediction'' and provide code to utilize analyses of and the development of ML applications on our data. All data and code is publicly available at \url{https://github.com/IllDepence/contextgraph}.
\end{abstract}

\ccsdesc[500]{Information systems~Information retrieval}
\ccsdesc[500]{Computing methodologies~Information extraction}
\ccsdesc[300]{Computing methodologies~Language resources}
\ccsdesc[500]{Computing methodologies~Knowledge representation and reasoning}
\ccsdesc[500]{Mathematics of computing~Graph theory}

\keywords{scholarly data, research artifacts, link prediction, information extraction}

\maketitle

\section{Introduction}
The rate of scholarly publication is consistently rising~\cite{Bornmann2021}, confronting researchers with an ever greater challenge of identifying research that is relevant to their endeavours and interests. To assist researchers in this task, several systems such as academic search engines and recommender systems, as well as prediction approaches have been developed.
However, existing approaches are limited in terms of the granularity and variety of entities that are subject to the search, recommendation, or prediction offered.
Many systems and methods focus on papers, for example, the recommendation of papers for reading based on a user profile.
Recent approaches operate on a more fine grained level---for example data set recommendation based on a research task description. These are, however, still constrained to a single type of entity (e.g. data sets only).
This is problematic because research endeavours commonly involve multiple ``research artifacts'' such as \emph{methodologies} and \emph{data sets}, which means single entity type methods only reflect them partially. Furthermore can the \emph{combined} use of existing ideas and artifacts be seen to reflect innovation, in academia~\cite{Hedlund2020,Devarakonda2020} as well as other collaborative areas~\cite{Tan2020}.

\begin{figure}
  \centering
  \includegraphics[width=\linewidth]{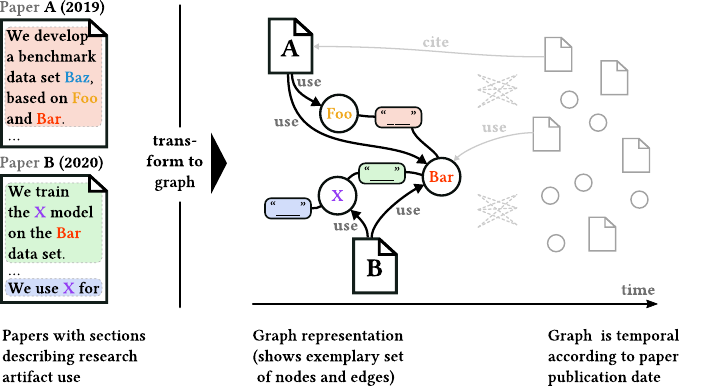}
  \caption{Schematic of our data set.\\
    \footnotesize\normalfont Created from a large corpus of paper full-texts (unarXive) and metadata on research artifacts (Papers With Code), our data set is a graph representing papers being published over time, and the research artifacts they use. Usage is contextualized through text sections within the papers.
  }
  \label{fig:schema}
\end{figure}

To enable the study of combined use of research artifacts, and to overcome aforementioned limitations, we propose CoCon, a data set capturing the combined use of research artifacts, contextualized through academic publication text. The proposed data set is a graph containing 35 k research artifacts as well as 340 k publications (graph nodes). As shown in Figure~\ref{fig:schema}, these are connected (graph edges) according to publications' \emph{use} of artifacts, as well as through a citation network. In addition to our data set, we provide a code base for easy extension to machine learning (ML) applications, as well as a formalization of a link prediction task in a dynamic graph for ``combined research artifact use prediction''. %

Overall, we make the following contributions.

\begin{itemize}
    \item We propose CoCon, a graph data set reflecting 35 k research artifacts and their contextualized use in 340 k academic publications. By providing CoCon we, enable the study of and development of ML applications for a more holistic picture of research endeavours than was possible before.
    \item We formalize a link prediction task for ``combined research artifact use prediction''.
    \item We provide code to facilitate easy updates of our data set as well as easy extension to additional ML tasks.
\end{itemize}

\section{Data Set}
We first discuss our process of data source selection and data scheme definition in accordance to each other. After that, we describe the resulting preprocessing steps as well as our final data set.

\subsection{Scope and Data Sources}
The term ``research artifact'' is used in the literature to refer to research outputs which can be re-used or further developed in subsequent work. Typical examples are datasets, software prototypes, ontologies, and methodologies~\cite{Nguyen2021,tsunokake2022}. 
We considered various data sources providing information on, content of, as well as associations between (1)~research artifacts and (2)~publications. In particular, the data sets SCIERC~\cite{luan2018scierc}, SciREX~\cite{Jain2020}, and Papers With Code\footnote{See \url{https://github.com/paperswithcode/paperswithcode-data}.} are candidates regarding research artifacts, while S2ORC~\cite{Lo2020} and unarXive~\cite{Saier2020unarXive} are viable data sources for publications.

For research artifacts, we decide to use Papers With Code in favor of SCIERC and SciREX. This choice is motivated by the fact SCIERC is rather small in comparison to the other two alternatives, and SciREX is built from Papers With Code. Directly using Papers With Code allows us to obtain a larger amount and also more recent data. Furthermore, it will allow us to update our data in the future, as Papers With Code periodically update their shared data.

Using Papers With Code as a data source for research artifacts has two important implications. First, the classes of research artifacts we are able to model are (1)~data set, (2)~method, (3) model, and (4)~task. Secondly, the field of publications covered is mainly from the area of machine learning. Because a majority of the publications in Papers With Code are linked to author copies on arXiv.org, both unarXive and S2ORC are a good fit, since both cover the entirety of arXiv papers. Between the two, we chose to use unarXive, which is motivated by two reasons. First, unarXive includes a more complete citation network (43\% compared to 31\%), and second, newer S2ORC releases do not include arXiv papers parsed from \LaTeX\ sources anymore\footnote{\textit{``Release: 2020-07-05 [...] omitted LaTeX parses from this release.''}, see \url{https://github.com/allenai/s2orc} (accessed 2023/02/12).}---which are less prone to noise than PDF parses.
Our data sources therefore are \textit{Papers With Code} and \textit{unarXive}. %

\subsection{Preprocessing}
The data provided by Papers With Code contains metadata on data sets, methods, models, and tasks, as well as the papers that use the aforementioned. However, the data model is not suitable for transformation into a graph \emph{as is}. In particular, not all entities are readily available as ``first class citizen'' records, but only described in relation to others. Models, for example, are only available as attributes of tasks. We transform the data accordingly, persisting (1)~for each artifact its set of metadata attributes provided by Papers With Code, and (2)~links from each artifact to the papers it is used by.

To extend our data with paper full-texts and subsequently usage contexts (see Figure~\ref{fig:schema}), we need to connect the paper metadata records in Papers With Code to unarXive. This process is straightforward by use of paper's arXiv ID. In addition to the paper full-texts, this also adds unarXive's citation network to our data.

For the extraction of paper text segments, in which the usage of an artifact is described, we match each occurrence of the artifact's name in all papers which Papers With Code lists as using the artifact. In most cases this is unproblematic, as artifact names are unique, capitalized proper names or acronyms. We note some exceptions of artifacts named ambiguously, such as the data sets ``seeds''\footnote{See \url{https://paperswithcode.com/dataset/seeds}.} and ``iris'',\footnote{See \url{https://paperswithcode.com/dataset/iris-1}.} which we filter heuristically to prevent noise.

\subsection{Data Set Description}

\begin{figure}
  \centering
  \includegraphics[width=\linewidth]{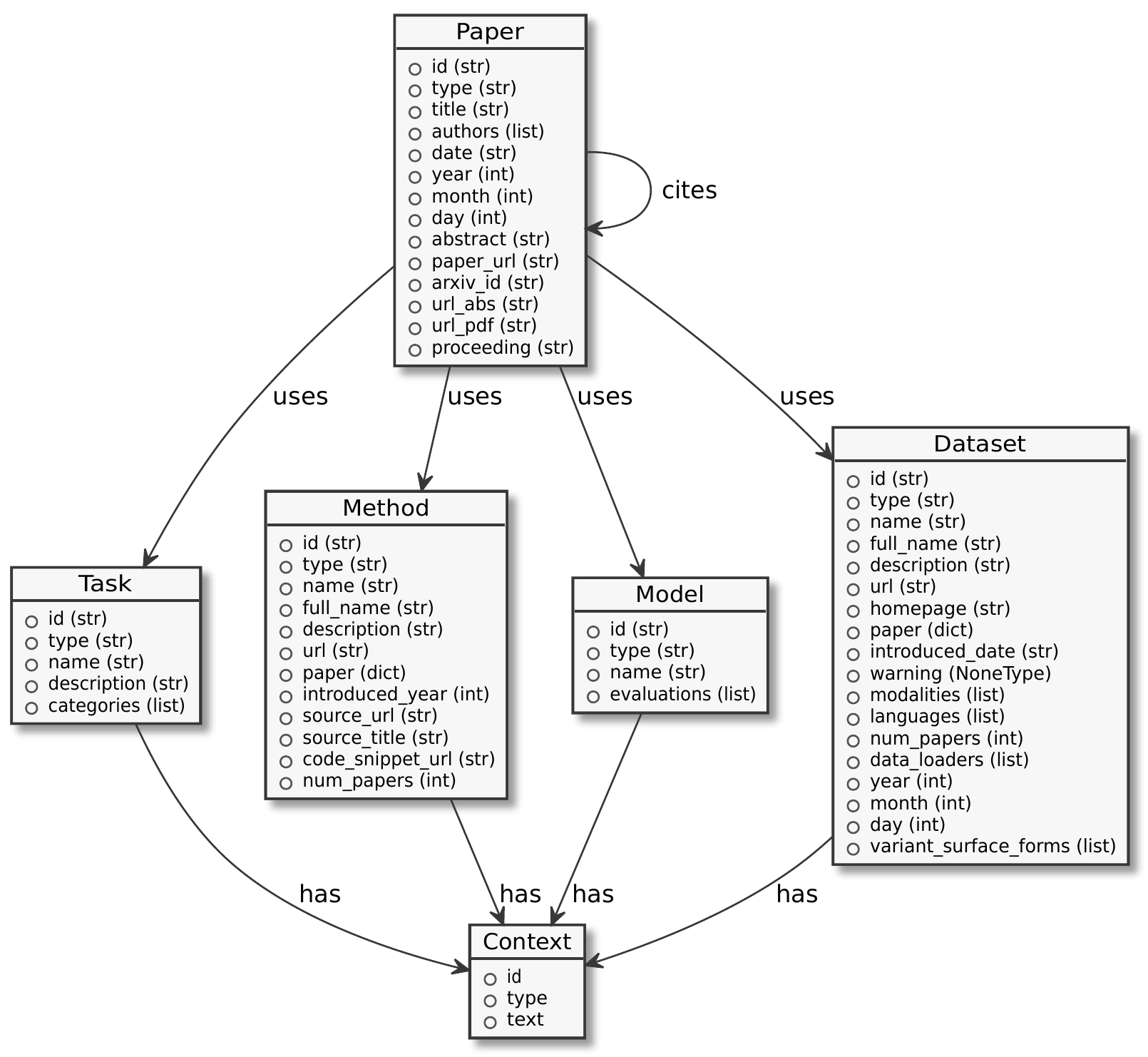}
  \caption{Graph schema of $\mathrm{CoCon}$.}
  \label{fig:graphschema}
\end{figure}

\begin{figure}
  \centering
  \includegraphics[width=\linewidth]{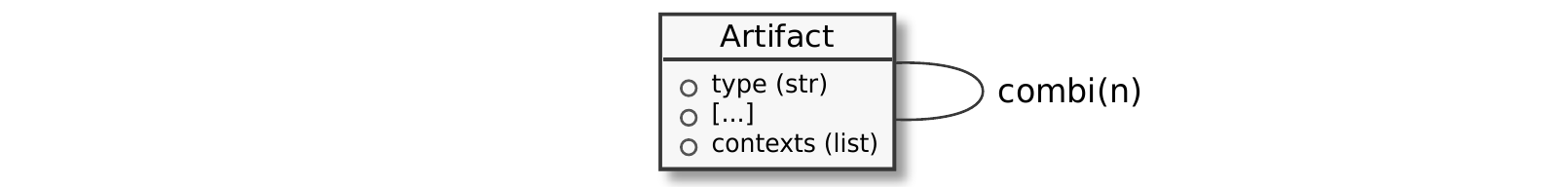}
  \caption{Simplified graph schema of $\mathrm{CoCon_{simple}}$.}
  \label{fig:graphschemasmol}
\end{figure}

Through the process described above, we get data on papers, artifacts, and their interconnections according to the schema shown in Figure~\ref{fig:graphschema}. We additionally define a simplified graph schema, $\mathrm{CoCon_{simple}}$, shown in Figure~\ref{fig:graphschemasmol}, where artifacts are connected by weighted edges according to the number of papers that use them together. Contexts in the simplified graph schema are represented as artifact nodes' attributes.

\begin{figure}
  \centering
  \includegraphics[width=\linewidth]{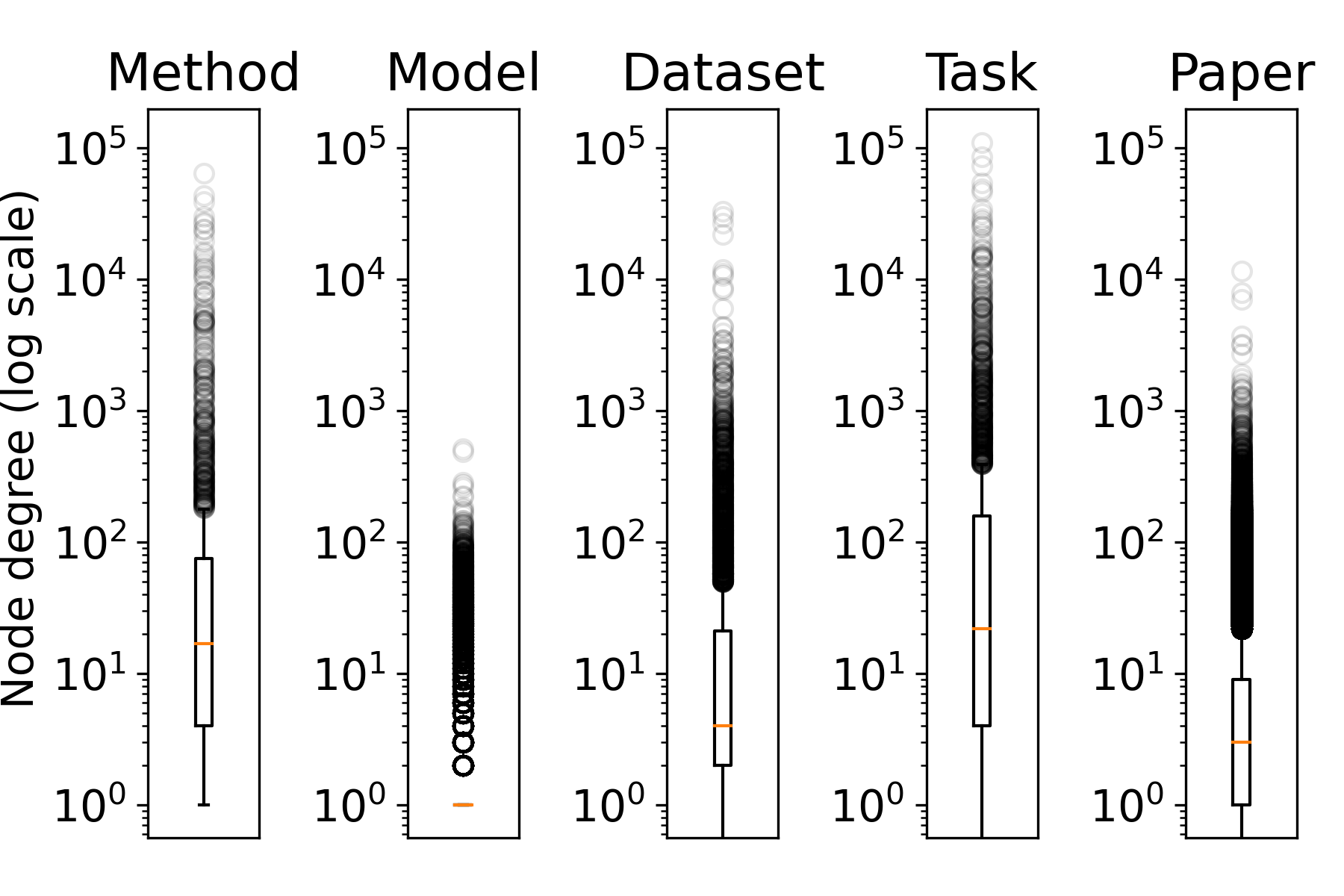}
  \caption{Node degrees.}
  \label{fig:node_degree}
\end{figure}

Our data set contains 340,965 paper nodes, 22,189 model nodes, 7,592 data set nodes, 3,389 task nodes, and 2,026 method nodes. Furthermore, it contains 2,413,664 contexts. Figure~\ref{fig:node_degree} shows the distribution of degrees across the different node types. We note a clear long tail distribution for all types---that is, most research artifacts and papers have few connections, while a few have have a large number of connections. A distinct case are models, most of which (72\%) have a degree of 1. This can be explained by the fact that authors use experiment specific model identifiers in tables describing evaluation results, which are the source of model names.\footnote{See \url{https://github.com/paperswithcode/sota-extractor}.}

To make out data set easily usable as a graph, we provide code to load it as a \texttt{networkx}\footnote{See \url{https://networkx.org/}.}, and \texttt{PyTorch Geometric}\footnote{See \url{https://pytorch-geometric.readthedocs.io/}.} graph, as shown in Listing~\ref{lst:pycode}.

\begin{lstlisting}[language=Python,caption=Python usage example.,label=lst:pycode,breaklines=true,captionpos=b,frame=single,showlines=true]

from contextgraph.util.graph import load_full_graph
G = load_full_graph(with_contexts=True)
print(type(G))
# output: networkx.classes.digraph.DiGraph

from contextgraph.util.torch import load_entity_combi_graph
G_combi = load_entity_combi_graph()
print(type(G_combi))
# output: torch_geometric.data.data.Data

\end{lstlisting}

\section{Link Prediction Task}\label{sec:background}

In line with existing work using the Papers With Code data, our data set can be used for various types of analyses and applications, such as the analysis of research artifact usage patterns~\cite{Koch2021}, the evaluation of information extraction approaches~\cite{Yousuf2022}, and content based paper recommendation~\cite{Ostendorff2022}.

As a novel application made possible by CoCon specifically, we formalize the link prediction task ``combined research artifact use prediction'' ($\mathrm{CRAUP}$). The reasons for choosing this task are twofold. First, while commonly used scholarly data sets for link prediction tasks exist, they typically consist only of paper nodes connected through citation edges and furthermore are comparatively small (see Cora, CiteSeer, and PubMed Diabetes in the CTU Prague Relational Learning Repository~\cite{ctu2015} %
used in, for example, \cite{Pan2018,Davidson2018,Ahn2021}). We therefore see a benefit in formalizing a link prediction task with a focus on research artifacts insofar, as a new type of real world phenomenon (use of research artifacts) is reflected. Secondly does the real world application scenario bear potential for the identification of beneficial research endeavours---for example, if reliable predictions can be made that a certain method is suited for application on a task, or that a certain data set is suited as training data for a model.
In the following, we present a brief description and formalization of the task.

\subsection{Task Description}\label{sec:taskdescription}

The goal of $\mathrm{CRAUP}$, as the name suggests, is to predict whether or not two research artifacts will be used together in the future. The basis for the prediction is the evolution of the graph reflecting publications and research artifacts. Papers within CoCon naturally have a publication date which, by extension, can be used to also date research artifacts---i.e., they enter the graph with the paper that first uses them. We can therefore view both our $\mathrm{CoCon}$ and $\mathrm{CoCon_{simple}}$ as a time discrete dynamic graph~\cite{Kazemi2022}, that is, a graph that changes in discrete time steps.

Formally, we can then define the prediction problem on the series of graph snapshots $G^{(0)}, ..., G^{(T)}$ with time steps $t \in T$. Each graph $G^{(t)} = (V^{(t)}, E^{(t)})$ consists of vertices $v \in V^{(t)}$ and edges $(u, v) \in E^{(t)}$. The types of possible changes from any $G^{(t)}$ to the subsequent $G^{(t+1)}$ are as follows.

\begin{itemize}
    \item Adding of nodes (papers published in $t+1$, and artifacts used for the first time in a papers published in $t+1$)
    \item Adding of edges (edges associated with above nodes)
    \item Increase of edge weights in $\mathrm{CoCon_{simple}}$ (the weight of an edge $(u, v)$ increases by the number of papers published in $t+1$ that use both $u$ and $v$)
\end{itemize}

\begin{figure}
  \centering
  \includegraphics[width=\linewidth]{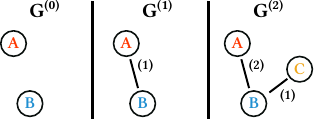}
  \caption{Visualisation of graph snapshots for $\mathrm{CoCon_{simple}}$.\\
    \footnotesize\normalfont At $G^{(0)}$, two papers are published, one of them using artifact A, the other one using artifact B. At $G^{(1)}$ a paper is published that uses artifacts A and B in combination. At $G^{(2)}$ a paper is published that uses artifacts A, B, and C in combination.\\
    \textbf{Note}: edge weights are indicated by numbers in brackets.
  }
  \label{fig:linkpred}
\end{figure}

In Figure~\ref{fig:linkpred} we show a simple visualisation, illustrating all possible changes over three graph snapshots in $\mathrm{CoCon_{simple}}$.

We define the most basic form of the prediction task, $\mathrm{CRAUP_{step}}$, on $\mathrm{CoCon_{simple}}$. The task is for every time step to predict which edges will be newly added to the graph. Formally, for each point in time $t \in T$ and each pair of nodes $u, v$ with $u \in V^{(t)}$, $v \in V^{(t)}$, and $(u, v) \notin E^{(t)}$, predict whether or not $(u, v) \in E^{(t+1)}$.

Because $\mathrm{CRAUP_{step}}$ is very specific in requiring an edge to be added at the immediate subsequent point in time $t+1$, we also define a more general task $\mathrm{CRAUP_{future}}$. Here, the prediction is not made for $(u, v) \in E^{(t+1)}$ but rather $(u, v) \in E^{(t+1, ..., T)}$, that is, predict whether or not the two artifacts $u$ and $v$ will be used together \emph{at any point} in the future, not necessarily already at the immediate subsequent point in time $t+1$.

As a further extension of $\mathrm{CRAUP_{future}}$, we define $\mathrm{CRAUP_{gap}}$, where for an additional time gap parameter $\Delta$ the prediction is made for $(u, v) \in E^{(t+1+\Delta, ..., T)}$. In other words, predict whether or not the two artifacts will be used together in the future, \emph{but not immediately}. Our reason for defining $\mathrm{CRAUP_{gap}}$ is to introduce a sense of ``early prediction''. The higher $\Delta$ can be set for a model without compromising on it's evaluation performance, the earlier it can predict whether or not two artifacts will be used together in the future.

We see investigations of the three tasks $\mathrm{CRAUP_{step}}$, $\mathrm{CRAUP_{future}}$, and $\mathrm{CRAUP_{app}}$ as a first step into exploring regularities in combined contextualized research artifact use over time. Further extensions to, for example, the inclusion of predicting edge weight changes in $\mathrm{CoCon_{simple}}$, link prediction tasks in $\mathrm{CoCon}$, or other types of prediction tasks, are potential following steps.

\section{Conclusion}\label{sec:conclusion}

We propose $\mathrm{CoCon}$, a large scholarly data set reflecting the combined use of research artifacts, contextualized in academic publications' full-text. The data set is a graph comprising 35 k artifacts (data sets, methods, models, and tasks) and 340 k publications, primarily from the area of machine learning. By focusing on artifact \emph{usage} as opposed to mentions or references in general, and by contextualizing usage in publications' full-text, we enable novel types of analyses and ML applications.

In addition to making $\mathrm{CoCon}$ freely available, we formalize a link prediction task ``combined research artifact use prediction'' ($\mathrm{CRAUP}$) and provide code to utilize analyses of and the development of ML applications with $\mathrm{CoCon}$. We plan to investigate $\mathrm{CRAUP}$ potential application scenarios of models developed in the process. This paves the way toward a new generation of systems aiding researchers in effectively identifying publications and research artifacts to use in their future endeavors.

\section*{Author Contributions}  %
Tarek Saier: Conceptualization (lead), Data curation, Formal analysis (lead), Methodology, Software, Visualization, Writing -- original draft, Writing -- review \& editing. Youxiang Dong: Conceptualization (support), Formal analysis (support). Michael F{\"a}rber: Writing -- review \& editing.

\begin{acks}
This work was partially supported by the German Federal Ministry of Education and Research (BMBF) via [KOM,BI], a Software Campus project (01IS17042)..
\end{acks}

\bibliographystyle{ACM-Reference-Format}
\bibliography{paper}

\end{document}